%% file: main-eurovis.tex
\documentclass{egpubl}
\usepackage{eurovis2020}

\usepackage{todonotes}
\usepackage{xspace}
\usepackage{xcolor,colortbl}
\usepackage{subfig}

\ConferenceSubmission   
\usepackage[T1]{fontenc}
\usepackage{dfadobe}  

\usepackage{cite}  
\BibtexOrBiblatex
\electronicVersion
\PrintedOrElectronic
\usepackage{caption}
\usepackage{caption}

\newcommand{\OurAlg}{ZMLT\xspace}

\usepackage{egweblnk}
\title{Multi-level tree based approach for interactive graph visualization with semantic zoom}
\author[]
{\parbox{\textwidth}{\centering
F. {De~Luca}$^{1}$\,
M.I. Hossain$^{1}$\,
K. Gray$^{1}$,
S. Kobourov$^{1}$\,
K. B\"{o}rner$^{2}$\
        }
        \\
{\parbox{\textwidth}{\centering 
$^1$University of Arizona, USA,
$^2$Indiana University, USA}}}

\begin{document}

 \teaser{
     \centering
   \includegraphics[height=1.8in]{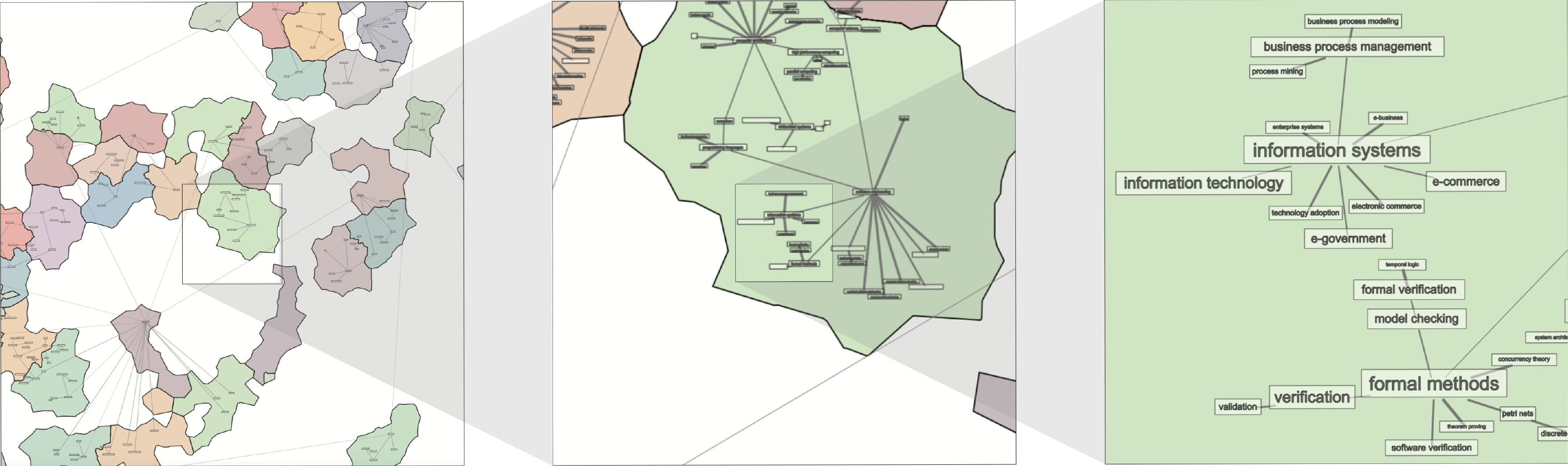}
   \caption{Map-based semantic-zoom graph visualization using a multi-level tree that is representative, real, persistent, overlap-free labeled, planar, and compact.}
   \label{fig:T8OurApproach}
}

\maketitle
\begin{abstract}
	Human subject studies that map-like visualizations are as good or better than standard node-link representations of graphs, in terms of task performance, memorization and recall of the underlying data, and engagement ~\cite{saket2014node,saket2015map}.
 	With this in mind, we propose the  \textit{Zoomable Multi-Level Tree} (\textit{ZMLT}) algorithm 
	for multi-level tree-based, map-like visualization of large graphs.  We propose seven desirable properties that such visualization should maintain and an algorithm that accomplishes them.
	(1) The abstract trees represent the underlying graph appropriately at different level of details; 
	(2) The embedded trees represent the underlying graph appropriately at different levels of details; 
	(3) At every level of detail we show real vertices and real paths from the underlying graph;
	(4) If any node or edge appears in a given level, then they also appear in all deeper levels;
	(5) All nodes at the current level and higher levels are labeled and there are no label overlaps;
	(6) There are no  edge crossings on any level;
	(7) The drawing area is proportional to the total area of the labels. 
	This algorithm is implemented and we have a functional prototype for the interactive interface in a web browser.
\end{abstract}

\section{Introduction}

\maketitle

Every day millions of people use maps to find restaurants, theaters, shops, and the best routes to reach them. 
GoogleMaps, AppleMaps, and OpenStreetMaps are used widely and they have trained many to read and use maps efficiently. In fact, most visitors to science museums can name and use maps effectively \cite{doi:10.1177/1473871615594652}. 
Online maps offer standard map interactions, such as zooming or panning:
panning moves the map in any direction while keeping it at the same scale and zooming changes the scale of the visualization by adding or removing details for a specific area (e.g., city, road, regional boundary demarcations and associated labels).
An important feature of most map visualizations is that when zooming in or out, the information density remains roughly the same and the type of information (e.g., major cities and roads) also remains the same. Cluster-based visualization approaches for large graphs, However, rely on meta-nodes (or super-nodes, cluster-nodes) and meta-edges (or edge bundling) instead of major ``nodes" and their key ``link" connections, and thus break the map metaphor. 
In order to take advantage of people's map reading skills when exploring graphs, we propose the  \textit{Zoomable Multi-Level Tree} (\textit{ZMLT}) algorithm for map-like visualization of large graphs. This algorithm offers one way to realize seven desirable properties of a map-based semantic multi-level tree representation of a large graph: 
\begin{enumerate}
    \item   \textbf{Appropriate  representation}: the abstract trees  represent the underlying graph appropriately at different levels of detail. 
    This property ensures that the graphs at each level of detail are connected (each level is a tree) and that the trees represent the underlying graph well on each level: Steiner trees~\cite{ahmed2018multilevel} are used to capture the most important nodes and edges.
    \item  \textbf{Appropriate layout}: the layout of the trees represents the underlying graph appropriately at different levels of detail. This property ensures that the global structure of the underlying graph is captured by the positions of the nodes in the layout, while simultaneously ensuring that the local structure is visible and readable. Further, important nodes have larger labels and important edges are thicker, as expected in multi-level map representations.
\item  \textbf{Real}: every node and every edge shown in every level of detail are ``real" as each level is a proper subgraph. This property prohibits meta-nodes and meta-edges, which would break the map metaphor, and is motivated by typical multi-level map representations.
\item  \textbf{Persistent}: if a node or edge appears at one level of detail, it will never disappear when zooming in further.
This property ensures persistence of the semantic level of detail representation, as expected in multi-level map representations. 
\item   \textbf{Overlap-free}: all nodes at the current and higher levels are labeled and there are no label overlaps. While theoretical graph layouts often consider each node to be a point, the visualization should take into account that each node has a different size labels that must be accommodated, as expected in multi-level map representations.
\item  \textbf{Crossing-free}: there are no edge crossings on any level, until the deepest level when the entire graph is shown.  Minimizing edge crossings is important in understanding graphs \cite{10.1007/3-540-63938-1_67}, and since we extract trees for each level, a layout without edge crossings is possible and desirable.
\item   \textbf{Compact}: the drawing area is proportional to the total area of the labels.  The sum of the sizes of each node label is the minimum area needed. However, this does not account for the space needed to show the edges.  A good visualization should have the labeled graph drawn in a compact way\cite{DBLP:journals/corr/abs-1801-07008}. 
This prevents the trivial solution of scaling the layout until all overlaps are removed, which creates vast empty spaces in the visualization. 
\end{enumerate}


\begin{figure}[tb]
	\centering
	\includegraphics[width=0.5\columnwidth]{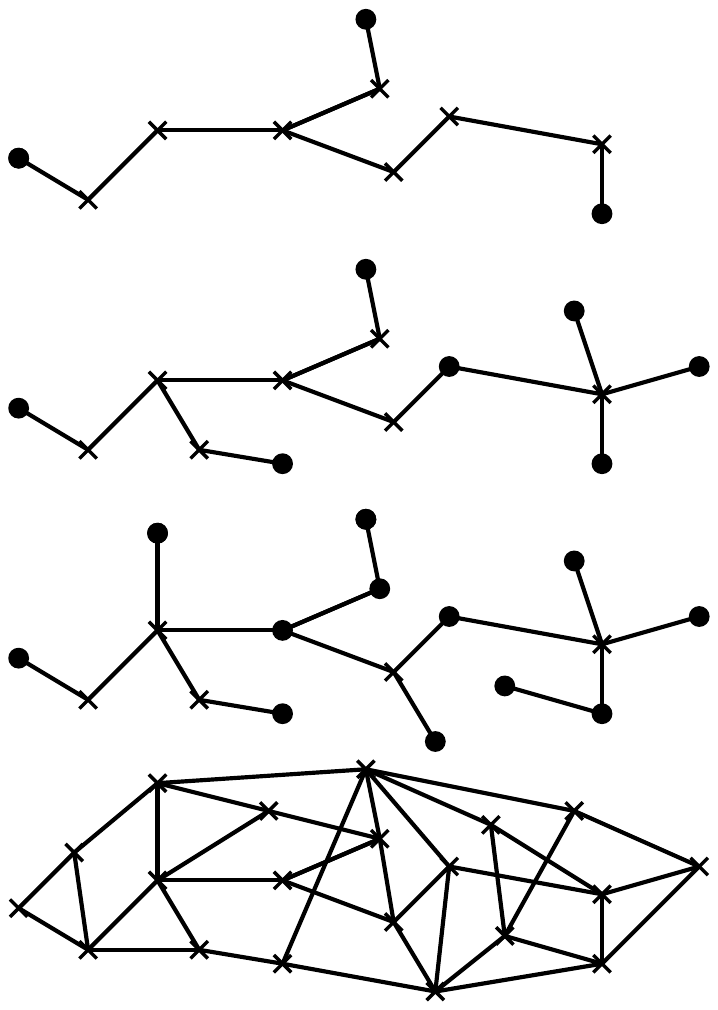}
	\caption{A multi-level Steiner tree: terminals are $\bullet$, Steiner points are $\times$}
	\label{fig:Steinertree}
\end{figure}

Formally, given a node-weighted and edge-weighted graph  $G=(V, E)$ we want to visualize $G$ as a set of progressively larger trees $T_1=(V_1, E_1) \subset T_2=(V_2, E_2) \subset \dots \subset T_n=(V_n, E_n) \subseteq  G$, such that $V_1 \subset V_2 \subset \dots \subset V_n = V$ and $E_1 \subset E_2 \subset \dots \subset E_n \subset E$. To make the trees representative of the underlying graph we rely on a multi-level variant of the Steiner tree problem~\cite{ahmed2018multilevel}, where we create the node filtration  $V_1 \subset V_2 \subset \dots \subset V_n = V$ with the most important nodes (highest weight) in $V_1$, the next most important nodes added to form $V_2$, etc. A solution to the multi-level Steiner tree problem then creates the set of progressively larger trees $T_1=(V_1, E_1) \subset T_2=(V_2, E_2) \subset \dots \subset T_n=(V_n, E_n) \subseteq  G$ using the most important (highest weight) edges; see~\autoref{fig:Steinertree}. Note that extracting the best tree $T_i$ that spans node set $V_i$ may require the use of nodes that are not in $V_i$, which are called Steiner nodes.

Just as in interactive map visualizations, at a minimum level of zoom, only the most important cities and highways are visible, and these are provided by the tree $T_1$ in our setting. Increasing the level of zoom brings in new cities and roads, and these are provided by the next tree in the filtration, until the maximum zoom level is reached and the whole map is shown (the entire underlying graph $G$, in our setting). In keeping with Tobler's first law of geometry\cite{10.2307/3693985}, our layout keeps similar nodes are closer to each other at each level of zoom. The seven guarantees above are embodied in the ZMLT layout, which is implemented and has a functional prototype for the interactive interface in a web browser; see \autoref{fig:T8OurApproach}.



The rest of this paper is organized as follows: In \autoref{sec:related} we describe related work, in \autoref{se:ourapproach} we give the details of ZMLT and in \autoref{se:visualizationtool} we describe the visualization tool to navigate and interact with the output of ZMLT. In \autoref{se:experiment} we compare the results of our approach to existing techniques. We conclude in \autoref{se:conclusions}.

\section{Related Work}~\label{sec:related}

\noindent \textbf{Large Graph Visualization:} Most graph layout algorithms use  a force-directed~\cite{eades1984heuristic,fruchterman1991graph} or a stress model~\cite{brandes2007eigensolver,koren2002ace} and provide a single static drawing.
The force-directed model works well for small graphs, but does not scale for large networks. Speedup techniques employ a multi-scale variant~\cite{hh-msadg-99,GGK04} or use parallel computation as in VxOrd~\cite{DWB01,bkk-mbs-05}. 
GraphViz~\cite{ellson2001graphviz} uses the force-directed method sfdp~\cite{hu2005efficient} that combines the multi-scale approach with a fast $n$-body simulation~\cite{Barnes86a}. 
Stress minimization was introduced in the more general setting of multidimensional scaling (MDS)~\cite{MDS} and has been used to draw graphs as early as 1980~\cite{seery1980designing}.
Simple stress functions can be efficiently optimized by exploiting fast algebraic operations such as majorization.
Modifications to the stress model include the  
strain model~\cite{torgerson1952},
PivotMDS~\cite{brandes2007eigensolver}, COAST~\cite{gansner2013coast}, and MaxEnt~\cite{gansner2013maxent}. 
Although these algorithms are fast
and produce good drawings of regular grid-like graphs, the results are not as good for dense, or small-world graphs~\cite{brandes2009}. 

Graph layout algorithms are provided in libraries such as GraphViz~\cite{ellson2001graphviz}, OGDF~\cite{chimani2011}, MSAGL~\cite{nachmanson2008}, and VTK~\cite{schroeder2000}, but they do not support interaction, navigation, and data manipulation. Visualization toolkits such as Prefuse~\cite{heer2005}, Tulip~\cite{auber2017tulip}, Gephi~\cite{bastian2009gephi}, and yEd~\cite{wiese2004yfiles} support visual graph manipulation, and while they can handle large graphs,
their rendering often does not: even for graphs with a few thousand nodes, the amount of information rendered statically on the screen makes the visualization unusable.

\noindent \textbf{Multi-Level Visualization:} Research on interactive multi-level interfaces for exploring large graphs inculdes ASK-GraphView~\cite{abello2006}, topological fisheye views~\cite{ky-vlgcfg-04,GKN05}, and Grokker~\cite{rivadeneira21}. Software applications such as Pajek~\cite{de2011} for social networks, and Cytoscape~\cite{shannon2003} for biological data provide limited support for multi-level network visualization. Zinsmanier \textit{et al.}~\cite{zinsmaier2012interactive} proposed a technique to deal with visual clutter by adjusting the level of detail shown, using node aggregation and edge bundling. 

\begin{figure}[t]
	\centering
	\includegraphics[width=1\columnwidth]{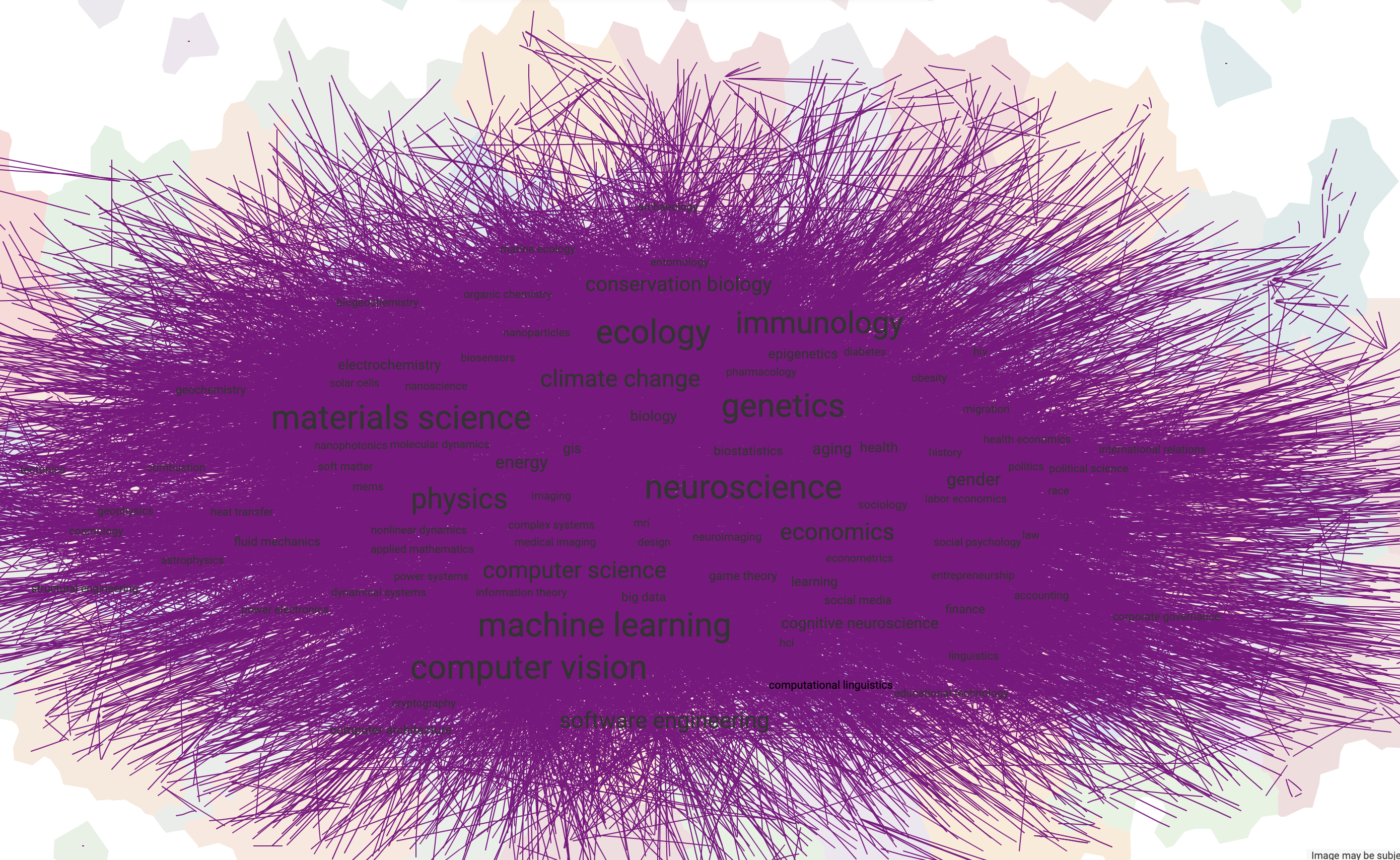}
	\includegraphics[width=1\columnwidth]{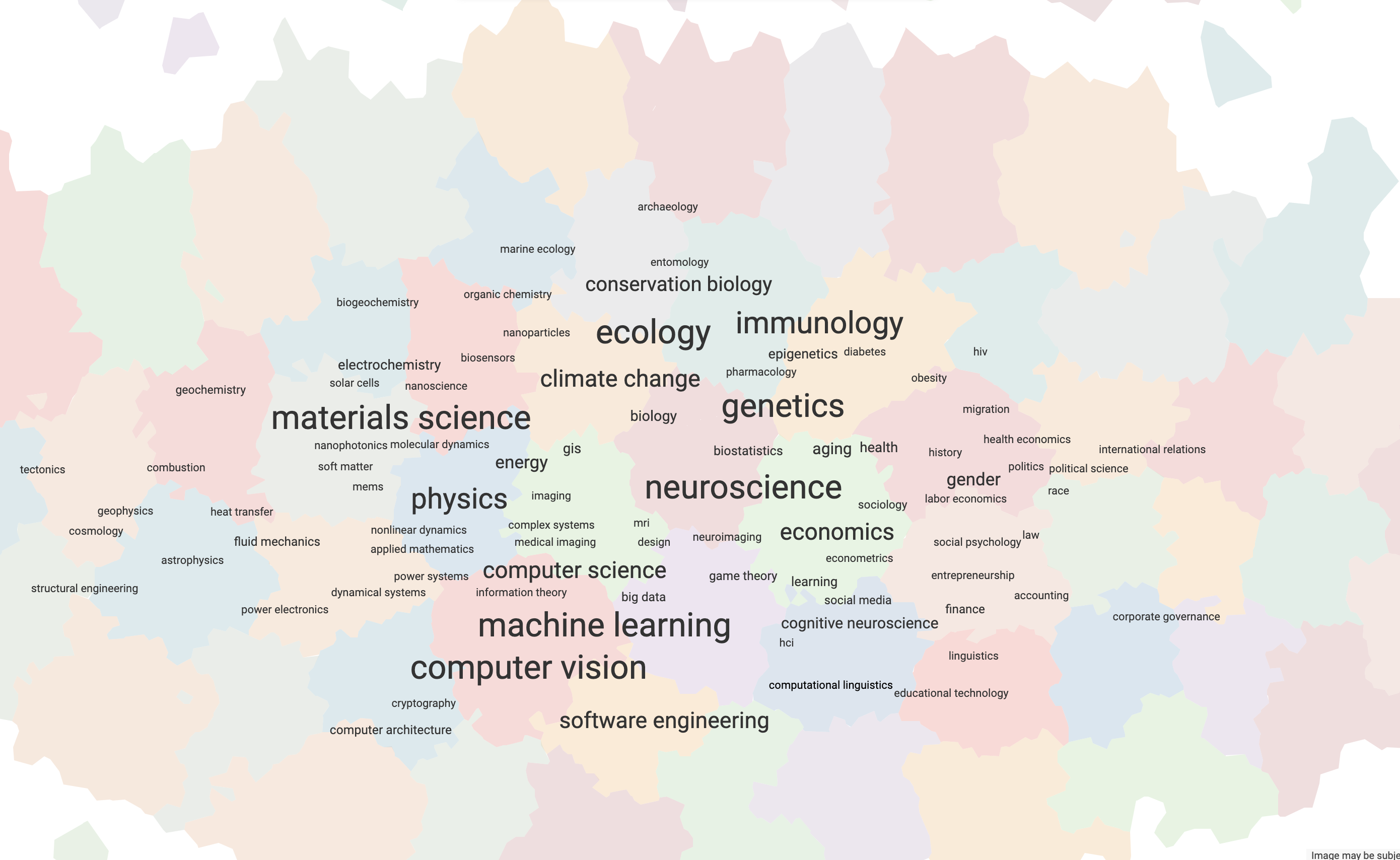}
	\caption{Screenshots from the GRAM~\cite{burd2018graph} system with and without edges.}
	\label{fig:Gramscreen}
\end{figure}

Most of these approaches rely on meta-nodes and meta-edges, which make interactions such as semantic zooming, searching, and navigation counter-intuitive,
as shown by Wojton \textit{et al.}~\cite{wojton2014sense}.
This work confirms that people have difficulty reading graph layouts that use hierarchical cluster representations (via high-level meta-nodes and meta-edges) such as ``super-noding" and ``edge bundling."
Other studies that compare standard node-link representations with map-like ones show that map-like visualizations are superior in terms of task performance, memorization and engagement~\cite{saket2014node,saket2015map}. 
In this context, \textit{GraphMaps}~\cite{modal2018new} visualizes a graph using the map metaphor and provides different zoom levels, driven by an embedding of the entire underlying graph.
GRAM~\cite{burd2018graph} also visualizes a graph using the map metaphor and provides different zoom levels but neither GraphMaps nor GRAM guarantees all seven of our desirable properties. 
For example, GRAM does not provide consistency or planarity. GRAM provides a button to show or hide the edges of the shown graph but the first option does not show the structure of the graph and the second one leads to a hairball; see \autoref{fig:Gramscreen}.

\noindent \textbf{Overlap Removal and Topology Preservation:} 
In theory, nodes are usually treated as point objects, but in practice, nodes are labeled and these labels must be shown in the layout. Overlapping labels is a major problem for most graph layout algorithms, and it severely affects the usability of the visualization. 
The standard approach to dealing with this problem is post-processing of the layout where node positions are perturbed to remove label overlaps. 
This usually results in significantly modified layout with increased stress, larger drawing area, and with the introduction of crossings even when the starting node configuration is crossings-free.

A simple solution to remove overlaps is  to scale the drawing until the labels no longer overlap. This approach works on every layout and is straightforward, although it may result in an exponential increase in the drawing area. 
Marriott \textit{et al.}~\cite{marriott2003removing} proposed to scale the layout using different scaling factors for the $x$ and $y$ coordinates. This reduces the overall blowup in size but may result in poor aspect ratio.
Gansner and North~\cite{gansner1998improved},  Gansner and Hu~\cite{gansner2009efficient}, and Nachmanson \textit{et al.}~\cite{nachmanson2017node} describe overlap-removal techniques with better aspect ratio and modest additional area. However, none of these approaches (except the straightforward scaling), can guarantee that no crossings are added when starting with a crossings-free input.

Post-processing a layout while preserving the topology of the input layout is a difficult task.  \textit{Pred}~\cite{bertault1999force} is a post-processing force directed algorithm applied on a given layout that aims to improve layout quality metrics while preserving the number of crossings. 
\textit{Impred}~\cite{simonetto2011impred} is an improved version that speeds-up the force computation and provides better results. This algorithm is not designed to remove the node overlaps but, with some appropriate modifications described in \autoref{se:removeoverlap}, it can be used for that purpose, while ensuring that the number of crossings remains unchanged.

\noindent \textbf{Steiner Trees:}
The  Steiner tree problem is one of Karp's  original 21 NP-Complete problems~\cite{karp1972}. A polynomial time approximation scheme exists if the underlying graph is planar~\cite{borradaile2007}, but not for general graphs~\cite{chlebik2008}.
More importantly, the Steiner tree can be efficiently approximated within a factor of $\ln 4 + e < 1.39$~\cite{byrka2010}. 
In the node-weighted Steiner tree problem, the cost of the tree is determined by the weights of the nodes  included in the solution (rather than the weights of the edges). This version of problem is also  NP-hard~\cite{moss2007} but can be approximated within a logarithmic factor~\cite{guha1999,klein1995nearly}. 
The classic Steiner tree problem has also been generalized to a multi-level setting~\cite{Balakrishnan1994ModelingAH,1288137,Chuzhoy:2008:AND:1361192.1361200}, where the terminals appear on different levels and must be connected by edges of appropriate levels, as described in the introduction. In particular, Ahmed \textit{et al}.~\cite{ahmed2018multilevel} showed that the standard (edge-weighted) Steiner tree problem can be efficiently approximated to within a constant factor. Similarly, Darabi \textit{et al.}~\cite{darabi2018approximation} described a multi-level version of the node-weighted Steiner tree, and showed that it can be approximated as well as the single-level version.

\section{The Zoomable Multi-Level Drawing Algorithm}\label{se:ourapproach}

\begin{figure}[htbp]
	\centering
	\subfloat[Orthogonal]{\label{fig:ortho}\includegraphics[width=.49\columnwidth]{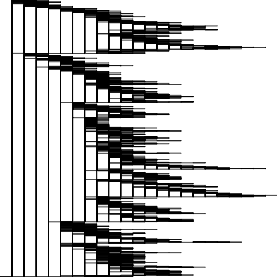}}
	\subfloat[Walker]{\label{fig:t8walker}\includegraphics[width=.49\columnwidth]{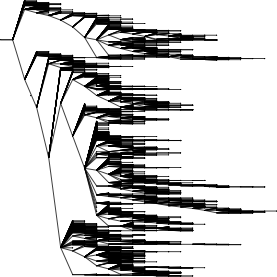}}\\
	\subfloat[Radial]{\label{fig:radial}\includegraphics[width=.48\columnwidth]{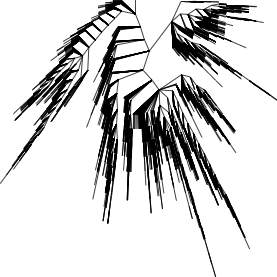}}
	\subfloat[Hierarchical]{\label{fig:hierarchical}\includegraphics[width=.49\columnwidth]{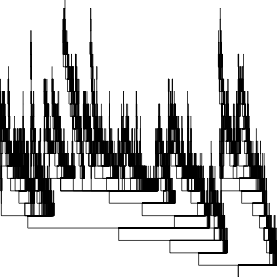}}
	\caption{Layouts of the large tree $T_8$ that is used as a running example in this paper, produced by several dedicated tree drawing algorithms in Tulip~\cite{auber2017tulip}.}
	\label{fig:walkerApproach}
\end{figure}

As discussed in the previous section, there are many graph drawing algorithms and systems, and there is a also a great deal of work on visualizing trees; see \textit{Treevis.net}~\cite{shulz2011treevis} which summarizes more than 300 techniques. However, none of these can guarantee the seven desirable layout properties that we need: appropriate representation, appropriate layout, real, persistent, overlap-free, crossing-free, and compact. For example,  sfdp~\cite{hu2005efficient} produces great drawings but cannot guarantee that trees are drawn without crossings. Other algorithms designed specifically for crossing-free tree layouts (e.g., Walker, orthogonal, radial, hierarchical) do not produce representative views of the underlying graph, as they impose a layer-by-layer visualization in graphs where there is no such structure. 
Further,  layer-by-layer visualizations may impose an implicit  readability direction of the drawing, which can be interpreted as capturing some notion of node importance based on placement in the layout; see \autoref{fig:walkerApproach}.


\begin{figure}[htbp]
	\centering
	\includegraphics[width=1\columnwidth]{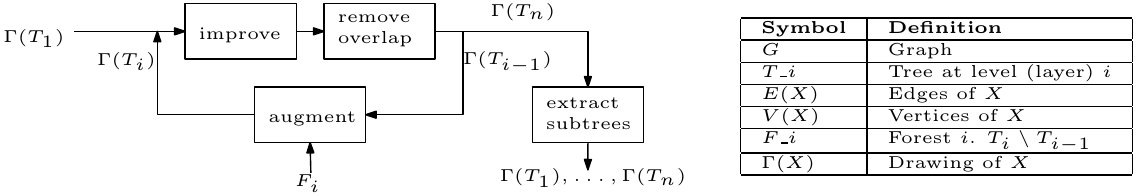}
	\caption{ZMLT framework to extract and compute layouts for the multi-level tree hierarchy (left) and a table of notation used (right).}
	\label{fig:framework_block}
\end{figure}

With this in mind, we designed the Zoomable Multi-Level (ZMLT) algorithm that guarantees all seven of the desired  properties. 
We begin with an overview of ZMLT as a block diagram shown in \autoref{fig:framework_block}.

The input of this algorithm is a weighted graph with non-negative weights on edges and nodes and the general framework consists of the following steps:
\begin{enumerate}
	\item \textit{Layered tree extraction}: Given weighted graph, extract a set of trees that are persistent, real and representative of the graph.
	\item \textit{First-layer drawing}: Draw the first (smallest) tree in a planar fashion.
	\item \textit{Improve}: Improve the quality of the produced drawing while maintaining planarity.
	\item \textit{Remove overlap}: Remove label overlaps while maintaing planarity and aiming for compactness.
	\item \textit{Augment}: Add to the current tree the forest of subtrees needed to obtain the next (larger) tree, while maintaining planarity.
	\item Repeat steps 3-5 until the last layer is reached.
	\item \textit{Extract subtrees}: Extract the layouts of all trees $T_1, T_2, \dots$ from the drawing of final tree that spans all the nodes of the graph. 
\end{enumerate}

\subsection{Layered tree extraction}
	\textbf{Input:} Weighted graph $G$.\\
	\textbf{Output:} Set of trees $\{T_1, \dots, T_n\}$ such that $T = T_1 \subset T_2 \subset \dots \subset T_n $.\\
	\textbf{Module:} Multi-level Steiner tree extraction algorithm.

This module deals with the construction of the (abstract) multi-level tree from the given weighted graph. 
We use a 
\textit{multi-level node-weighted Steiner tree (MVST)} ~\cite{darabi2018approximation}, defined and discussed in the Introduction. 
For our purposes we are interested in the maximum weight Steiner tree, as nodes and edges with high weights are important and should be present on the high levels in the multi-level hierarchy. 
While the Steiner tree problem usually asks for the minimum cost tree, this is easy to fix by simply choosing the reciprocal of the weights (assuming all weights are positive and greater than 1). 
Note that the multi-level Steiner tree step above results in the creation of abstract (not yet drawn) trees. Before starting the actual layout computation, we pre-process the nodes and the edges. First, we assign different font sizes to the node labels, depending on the level they belong to. 
This approach can be extended to unweighted graphs, where node weights can be assigned based on structural graph importance, e.g., node degrees, or  betweenness centrality.

\paragraph{Rationale:} The multi-level Steiner tree  approach used in this module ensures the \emph{ appropriate representation} property, as important nodes and edges are present in proper levels. Further the multi-level Steiner tree approach guarantees the \emph{ real} property as real nodes and real paths are represented in every tree, no nodes or edges are created. Finally, this approach also guarantees the \emph{ persistent} property, as by construction we have that  $T_1 \subset T_2 \subset \dots \subset T_n$.

Differentiating the font sizes helps identify the more important nodes even on deeper levels when there is a mix of nodes from different levels. We associate an enclosing rectangle with each label and these are used to compute and remove the overlap in later stages. We ignore the weights from this point onward since they already played their role in the extraction of the multi-level Steiner tree. 

\subsection{
	First-layer drawing}\label{sec:impred}
	\textbf{Input:}  $T_1$.\\
	\textbf{Output:} Planar $\Gamma(T_1)$.\\
	\textbf{Module:} Crossing free drawing algorithm.

The input at this step is the tree at the first layer of the hierarchy, namely $T_1$. Here the desired output is a planar, crossing-free drawing of the unweighted, unlabeled tree. 
Algorithms that guarantee a planar drawing of $T_1$ and do not impose additional features (such as hierarchical visualization) are suitable for this module. 

\paragraph{Rationale: } 
Here we aim to compute an \emph{ appropriate representation} and \emph{ appropriate layout} for $T_1$ while ensuring the \emph{ planar} property.

\subsection{Drawing-improvement}
	\textbf{Input:} $\Gamma(T_i)$.\\
	\textbf{Output:} Improved $\Gamma(T_i)$.\\
	\textbf{Module:} Modified \textit{ImPred} algorithm.

This module improves the drawing of $\Gamma(T_i)$, the output of the previous step. 
The algorithm used for this module is \textit{ImPred}~\cite{simonetto2011impred}, a force directed algorithm that aims to improve the layout of a drawing without changing the number of crossings in the input layout.
To preserve the number of crossings of the input drawing, this algorithm uses three forces: ($i$) node-to-node repulsion force ($\textbf{F}^r_v$), ($ii$) edge attraction force ($\textbf{F}^a_v$), ($iii$) node-to-edge repulsion force ($\textbf{F}^e_v$). The last force prevents a node from crossing any edge and ensures that no crossing is removed or added.
%
We modified the desired edge-length distances in order to have longer edges at higher level trees and shorter edges for lower level trees. 

 Since in this step ImPred is applied on a tree, for each node $v$ and edge $u$ the algorithm computes the force $\textbf{F}^e_v$; this has an impact on the  scalability of the algorithm.
 
\paragraph{Rationale: }
This step is particularly meaningful from $T_2$ and deeper, when an already drawn high-level tree $T_i$ is combined with a forest of subtrees from the next level to obtain $T_{i+1}$. As we will see in Step 3.5 (Augment layer), the forest of subtrees is drawn in a sub-optimal way in order to ensure that it can be added while guaranteeing planarity. This makes the drawing improvement stage essential in the multi-level layout pipeline.
 Although this step may be time consuming, it is part of a pre-proccesing procedure and thus it does not affect the visualization and interaction with the final output.

\subsection{Remove label overlap}\label{se:removeoverlap}
	\textbf{Input:} $\Gamma(T_i)$.\\
	\textbf{Output:} $\Gamma(T_i)$.\\
	\textbf{Module:} Modified ImPred algorithm in \textit{remove overlap mode}.

In this module we add labels and remove overlaps. We do this by adding another \textit{Label-to-Label-repulsion (llrep)} force to ImPred. As in other force-directed algorithms, this force uses a repulsion force parameter $\delta$.  At each iteration we compute $llrep_{uv}$ for every pair of nodes $u$ and $v$ as follows.
Let $(x_u, y_u)$ and $2w_u, 2h_u$ be the coordinate, box width, and height of a node $u$, respectively. 
Let $(x'_u, y'_u)$ be the new force of $u$. This repulsion force is zero when nodes $u$ and $v$ do not overlap each other i.e.,  $w_u + w_v - abs(x_u - x_v) <0$ or $h_u + h_v - abs(y_u - y_v)<0 $. Otherwise the force is  applied in the direction of the line through the nodes $(x'_u, y'_u)=  
(-\delta, s * (x_u-\delta-x_v))$ and $(x'_v, y'_v)=  
(\delta, s *  \delta + y_u-y_v)$, here $s=\frac{y_v - y_u}{ x_v - x_u}$. 
Note that this approach reduces label overlap iteratively. In our experiments all overlap can be removed, but we also post-process by scaling in order to guarantee that there are no overlapping labels.


\paragraph{Rationale:} ImPred, with the additional Label-to-Label-repulsion force, works best locally. Thus the progressive multi-layer drawing algorithm has a good chance of removing overlaps without needing to resort to scaling. 

\subsection{Augment-layer}
	\textbf{Input:} $\Gamma(T_i)$, $F_{i+1}$.\\
	\textbf{Output:} Preliminary drawing of tree $T_{i+1}$.\\
	\textbf{Module:} Monotone drawing algorithm to add the forest.

\begin{figure}[htbp]
	\centering
	\includegraphics[width=\columnwidth]{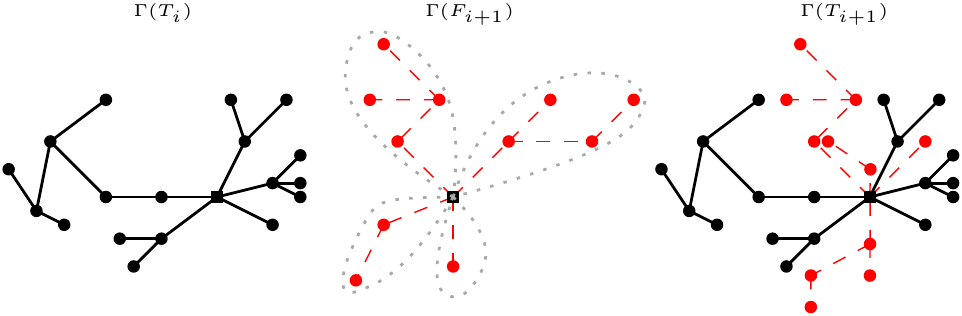}
	\caption{Augmentation: $\Gamma (T_i)$ is augmented with $\Gamma (F_{i+1})$ to obtain $\Gamma (T_{i+1})$. In this example, the common node of $T_i$ and $F_{i+1}$ is shown as a square.
		Each subtree rooted in the common node (enclosed by dotted gray region) is drawn in monotone way and placed in the cones of $\Gamma (T_{i+1})$ around the common node, based on the size and width of the cones. If crossings occur the subtree is scaled down until there are no crossings.}
	\label{fig:augmentation}
\end{figure}

This step adds to $T_i$ the nodes and edges of the forest $F_{i+1}$, so that $T_i\cup F_{i+1}=T_{i+1}$.
The goal of this module is to start with a good layout of $T_i$ and initial layouts of the forest $F_{i+1}$ and create a planar drawing of $T_{i+1}$. 

A monotone drawing of a tree is a straight-line drawing such that, for every pair of nodes, there exists a path that monotonically increases with respect to some direction~\cite{angelini2011monotone}. Since a monotone drawing of a tree is always a planar drawing, we use it to add the drawing of $\Gamma(F_{i+1})$ to $\Gamma(T_i)$ yielding $\Gamma(T_{i+1})$.

Forest $F_{i+1}$ is made of trees rooted in nodes of  $T_i$ and those are the only nodes in common between $F_{i+1}$ and $T_{i}$. The common nodes define where the rooted trees should be placed in the input drawing.
Each such subtree is placed in the middle of the cone defined by two consecutive edges around the common node, in decreasing order by size of the tree and the size of the cone. 
After placing the monotone drawing of the subtree a crossing check is performed. If a crossing is generated by this placement, the subtree is scaled down. This process continues until adding the subtree creates no crossings and until all components of the forests are placed; see \autoref{fig:augmentation}.

\paragraph{Rationale:} The main objective here is to combine $T_i$ with $F_{i+1}$ and obtain a planar drawing of $T_{i+1}$. The improvement of the layout and overlap removal are performed in other steps.

\subsection{Extract subtrees }
	\textbf{Input:} $\Gamma(T_n)$, all edges. \\
	\textbf{Output:} $\Gamma(T_1), \dots, \Gamma(T_n), \Gamma(G)$.\\
	\textbf{Module:} Adding edges.
This step actually computes the drawings for all the trees in the multi-level hierarchy. We simply extract from $\Gamma(T_n)$ the layouts for all subtrees $\Gamma(T_1), \dots, \Gamma(T_n)$. We also generate the layout for the entire graph, $\Gamma(G)$, by adding to $T_n$ all the edges of $G$.

\section{Interactive 
Interface and Prototype}\label{se:visualizationtool}
In this section we describe a functional browser interface and a prototype deployment of ZMLT with a specific dataset. 

\subsection{Web Browser Interface}
The output of the algorithm contains the positions of all the nodes in all the trees, as well as in the underlying graph. We modify the node-link representation into a map-like one using \textit{GMap}~\cite{gansner2010gmap}. Finally we extract GeoJSON files and use openlayers to display the data in the browser. 
We briefly describe the process.

\noindent \textbf{Node Clustering:} By default, we use MapSets~\cite{kobourov2014visualizing} to cluster the graph and show the clusters as contiguous regions on the map, although other GMap~\cite{gansner2010gmap} options are also available. 

\noindent \textbf{Generating Map Layers:} GeoJSON~\cite{butler2008geojson} is a standard format for representing simple geographical features, along with their non-spatial attributes. We split all geometric elements of the map into  \textit{ nodelayer},  \textit{ edgelayer}, and  \textit{ clusterlayer}. Each of the map layers is composed of many attributes. For example, nodelayer contains  node-ID, coordinates, font-name, font-size, height, width, label and weights for all nodes.

\begin{figure}[ht]
	\centering
	\includegraphics[width=1\columnwidth]{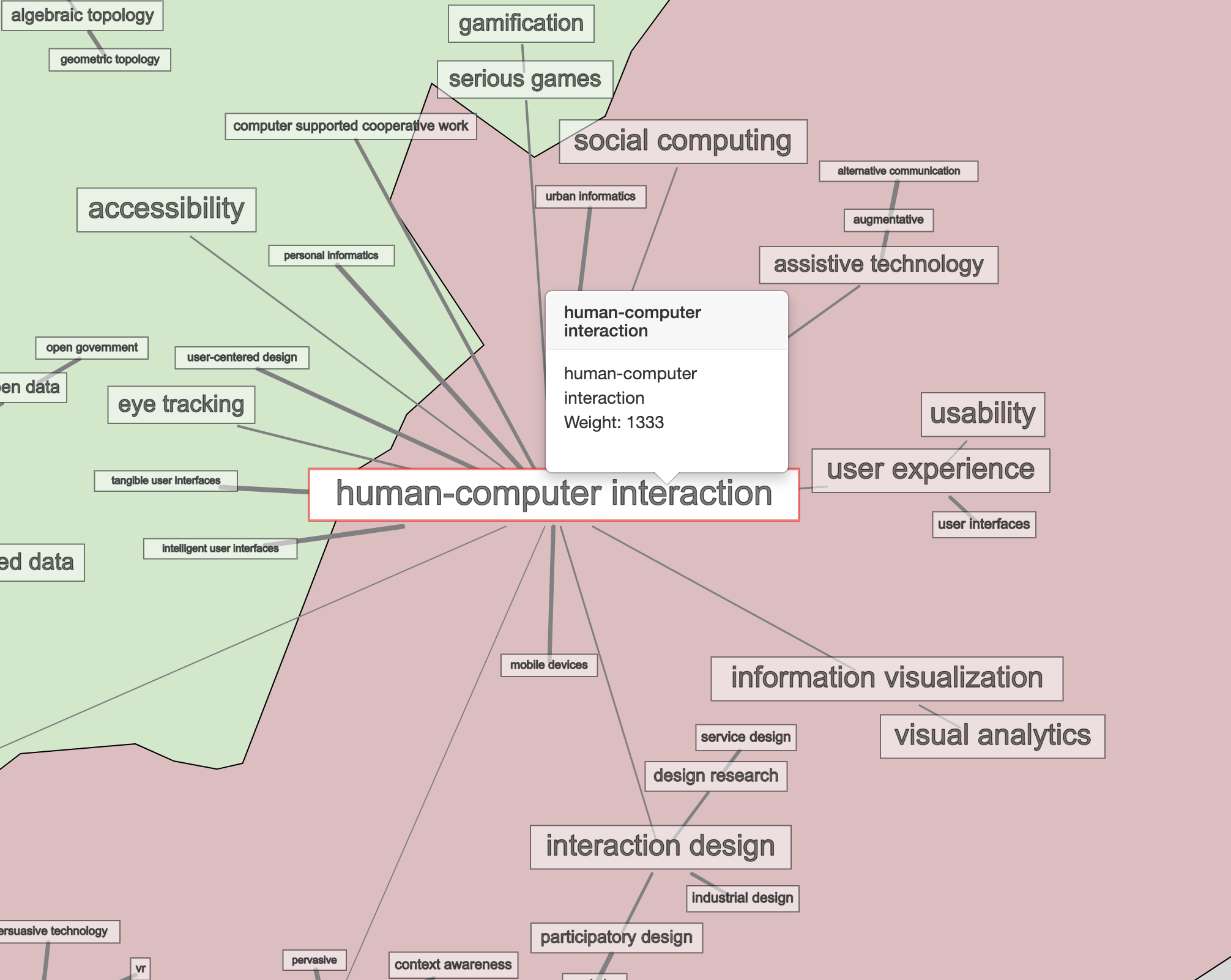}\caption{Node click in ZMLT visualization.}
	\label{fig:click}
\end{figure}
\noindent \textbf{JavaScript Application:} For rendering and visualizing the map layers we use openlayers, a JavaScript library for displaying map data in the browser. This visualization requires additional work to show the map elements in a multi-layer fashion. Node attributes, edge attributes and zoom level is used to determine the appropriate levels in the multi-level visualization.

\noindent \textbf{Interactions:} As an interactive visualization tool, we enable panning and zooming, as well as clicking on nodes and edges to see details such as weight; see the \autoref{fig:click}.

\subsection{Prototype ZMLT Deployment}\label{se:experiment}
In this section, we show the result of our technique applied to the GRAM Topics network~\cite{burd2018graph}, which is a weighted topic graph extracted from Google Scholar and contains $5947$ nodes and $26695$ edges.

 
 
We used the ZMLT algorithm with 8 levels, determined by node weights. Tree $T_1$ contains the heaviest $5\%$ of the nodes, tree $T_2$ contains the heaviest $15\%$ of the nodes, and so on until tree $T_8$ which contains all nodes. The exact percentages used are $(5, 15, 30, 40, 60, 70, 85, 100).$
All nodes are labeled using 8 different font sizes corresponding to their level.
The nodes in tree $T_1$ have font size 30, the nodes in tree $T_2$ have font size 25, and so on until tree $T_8$ where nodes have font size 8. The exact font sizes are $(30, 25, 20, 15, 12, 10, 9, 8)$. 


The prototype is available here: \url{http://uamap-dev.arl.arizona.edu:8086/}.

\subsection{Evaluation}
We compare ZMLT to a simpler implementation relying on off-the-shelf tools using several quality metrics, on the graph mentioned above: the GRAM Topics network.  We go over each of the seven properties next.

\begin{enumerate}
    \item The abstract trees are created using an algorithm for computing multi-level max Steiner tree. The underlying problem is NP-hard, but we use a constant approximation algorithm which has been shown to produce near optimal solutions in practice. Thus, the multi-level Steiner tree represents the underlying graph well. 
    
    \item The embedded trees provide an appropriate representation. We evaluate this quantitatively by measuring the stress in the layout, which captures how well the embedding preserves graph distances. Further, we measure multi-level edge uniformity which captures how well the the embedding captures the multi-level hierarchy.  
    
    \item Real nodes and paths are shown.  This property is ensured by the algorithms, as we never create or merge nodes or edges.
    
    \item Any node or edge that appears on a higher level also appears on the lower levels.  This property is ensured by the algorithm as the multi-level Steiner trees provide persistence.
    
    \item There are no edge crossings.  This property is ensured by the algorithm at every level.
    
    \item The drawing area is compact and we evaluate this quantitatively using the compactness measure.
\end{enumerate}


\noindent \textbf{Direct Approach:}
Our goals of ensuring a crossing-free output, or planarity, and labeling all nodes at the current and higher levels without overlaps  can be achieved using the  off-the-shelf \textit{Circular Layout} offered by \textit{yED}. This algorithm produces layouts that emphasize group and tree structures within a network drawing trees in a radial tree layout fashion~\cite{wiese2004yfiles}. 
We produced this layout by drawing the last layers of the hirarchies, namely, $T_8$ using Circular Layout with BCC Compact and the ``consider node labels" feature. Then we extracted the subtrees of the hierarchy.
In the following we call this procedure \textit{Direct Approach}. 

\noindent \textbf{Quality Metrics:}
ZMLT uses the graph weights only for the extraction of the trees. As both the direct approach and ZMLT draw unweighted trees, we then use the following three (unweighted) metrics:
\begin{itemize}
	\item
	Desired edge length (\texttt{DL}) evaluates the normalized desired edge lengths in the each layers. Since higher levels include the most important nodes and edges they correspond to larger vertices and longer edges;

	\item Stress (\texttt{ST}) captures how well the geometric distances in the layout between pairs of nodes reflect their graph-theoretic distances in the graph;
	\item Compactness (\texttt{CM}) measures the ratio between the sum of the areas of labels and the area of the actual drawing.  We consider that the minimum area to show all labels without overlap is the sum of their area. We then report the area amount of area that is used by the drawing algorithm as the ratio of the area of the drawing and the area of the labels. Since the sum of the areas of the labels is constant this measure can be used to compare areas of different layouts.
\end{itemize}

All metrics are available at \url{https://github.com/felicedeluca/graphmetrics}. 

\begin{figure}[thp]
	\centering
	\subfloat[$T_1$]{\label{fig:t1direct}\includegraphics[width=.24\columnwidth]{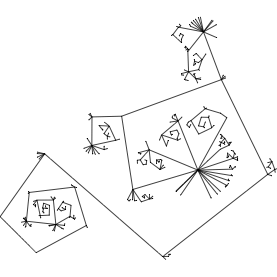}}
	\subfloat[$T_2$]{\label{fig:t2direct}\includegraphics[width=.24\columnwidth]{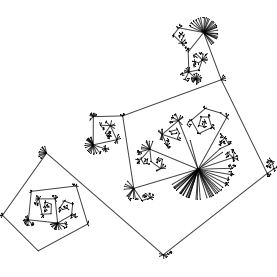}}
	\subfloat[$T_3$]{\label{fig:t3direct}\includegraphics[width=.24\columnwidth]{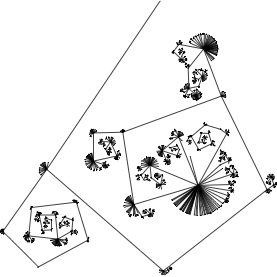}}
	\subfloat[$T_4$]{\label{fig:t4direct}\includegraphics[width=.24\columnwidth]{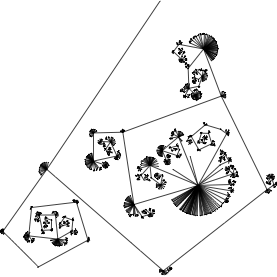}}\\
	\subfloat[$T_5$]{\label{fig:t5direct}\includegraphics[width=.24\columnwidth]{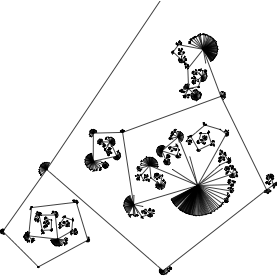}}
	\subfloat[$T_6$]{\label{fig:t6direct}\includegraphics[width=.24\columnwidth]{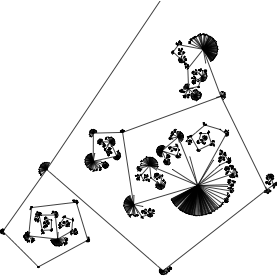}}
	\subfloat[$T_7$]{\label{fig:t7direct}\includegraphics[width=.24\columnwidth]{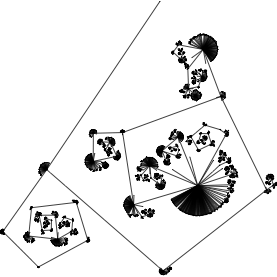}}
	\subfloat[$T_8$]{\label{fig:t8direct}\includegraphics[width=.24\columnwidth]{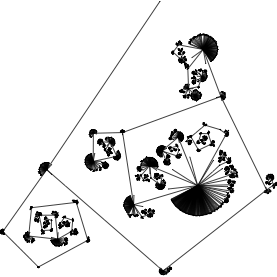}}
	\caption{Overview of the tree hierarchy structure of the Google Topic graph drawn with the Direct Approach.}
	\label{fig:DirectApproach}
	\subfloat[$T_1$]{\label{fig:t1impred}\includegraphics[width=.24\columnwidth]{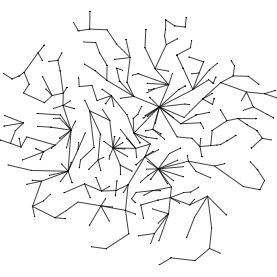}}
	\subfloat[$T_2$]{\label{fig:t2impred}\includegraphics[width=.24\columnwidth]{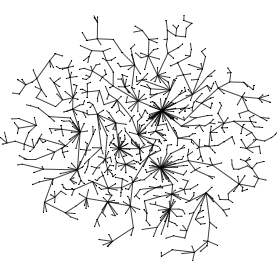}}
	\subfloat[$T_3$]{\label{fig:t3impred}\includegraphics[width=.24\columnwidth]{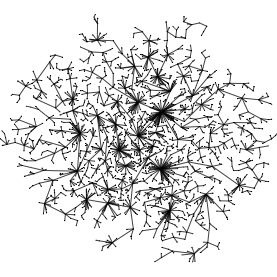}}
	\subfloat[$T_4$]{\label{fig:t4impred}\includegraphics[width=.24\columnwidth]{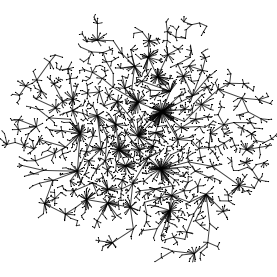}}\\
	\subfloat[$T_5$]{\label{fig:t5impred}\includegraphics[width=.24\columnwidth]{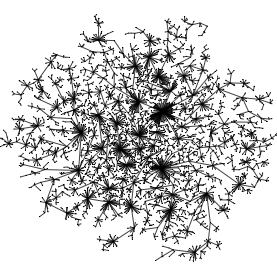}}
	\subfloat[$T_6$]{\label{fig:t6impred}\includegraphics[width=.24\columnwidth]{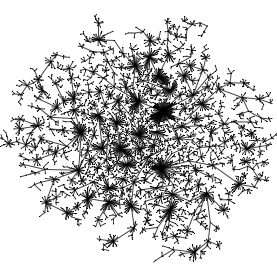}}
	\subfloat[$T_7$]{\label{fig:t7impred}\includegraphics[width=.24\columnwidth]{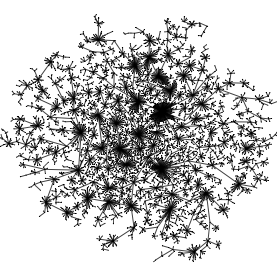}}
	\subfloat[$T_8$]{\label{fig:t8impred}\includegraphics[width=.24\columnwidth]{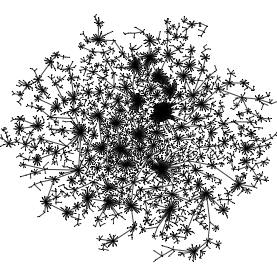}}
	\caption{Overview of the tree hierarchy structure of the Google Topic graph drawn with ZMLT.}
	\label{fig:OurApproach}
\end{figure}

\noindent \textbf{Results for the Google Topics graph:}
 An overview of the produced hierarchies are given by 
\autoref{fig:DirectApproach}  that shows the layouts of each layer computed by the direct approach  and by \autoref{fig:OurApproach}  that shows the layouts produced by our technique. Even though details are impossible to see,  this overview, regardless the correctness of the layout, highlights some significant differences stand out visually. For example, in order to achieve compact layout, the direct approach wraps the trees around in a spiral fashion, which is not a good representation for the underlying graph. There are also very large differences in the lengths of the edges produced by the direct approach.  The two overview give at a glance an information about the compactness of the structure of the two produced layout. In fact, the layout inf \autoref{fig:OurApproach} shows less blank space than \autoref{fig:DirectApproach} that is the layout produced by our approach is more compact.

\begin{figure}[htbp]
	\centering
	\includegraphics[width=1\columnwidth]{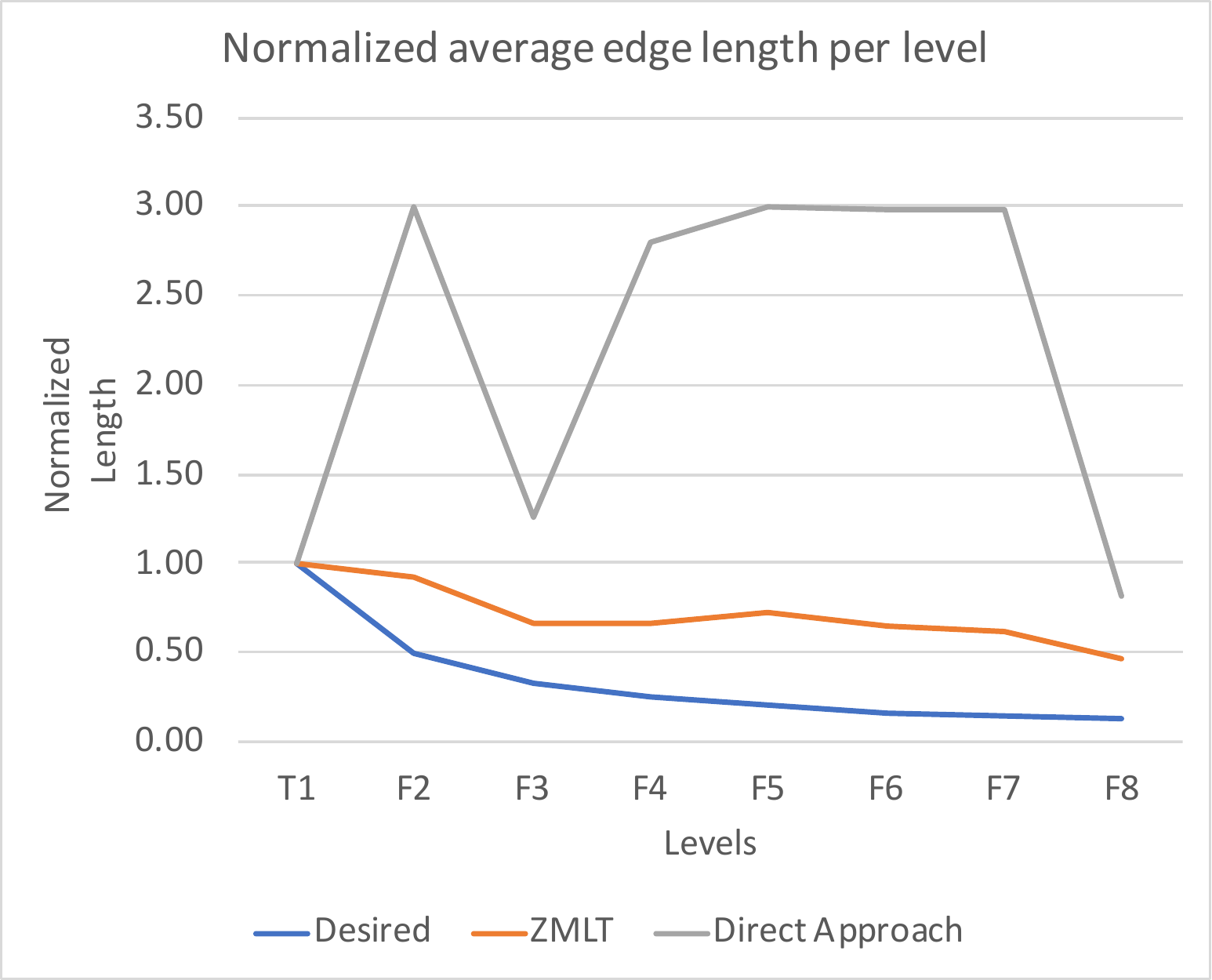}
	\caption{Trends of the edge lengths of the hierarchy: desired (in blue), computed by \OurAlg in orange, and computed by direct approach (in gray). These values are normalized to the average edge length of the first level.}
	\label{fig:alllengthscharts}
\end{figure}

The metric analysis in \autoref{tab:tabres} confirms the observations above. The table reports the values for the three measures computed on each of the 8 trees (representing the graph on different levels of detail). 
Low values in all metrics (ST, CM) correspond to better results and ZMLT has lower values in every table entry. Our approach has better performance with respect to the DL metric as also shown in \autoref{fig:alllengthscharts}.

\begin{table}[]
	\centering
	\parbox{.45\columnwidth}{
		\centering
		\small
		\begin{tabular}{|l|l|l|l|l|}
			\hline
			~&\texttt{DL}&\texttt{ST}&\texttt{CM}\\ \hline
			\textbf{$T_1$}&1&15502&9670\\ \hline 
			\textbf{$T_2$}&3.00&123164&4130\\ \hline 
			\textbf{$T_3$}&1.25&479681&2860\\ \hline 
			\textbf{$T_4$}&2.79&851388&2480\\ \hline 
			\textbf{$T_5$}&2.99&1894046&2100\\ \hline 
			\textbf{$T_6$}&2.98&2552986&1980\\ \hline 
			\textbf{$T_7$}&2.98&3736676&1840\\ \hline
			\textbf{$T_8$}&0.82&5138014&1740\\ \hline 
		\end{tabular}
	}
	\hfill
	\parbox{.45\columnwidth}{
		\centering
		\small
		\begin{tabular}{|l|l|l|l|}
			\hline
			~&\texttt{DL}&\texttt{ST}&\texttt{CM}\\ \hline 
			\textbf{$T_1$}&1&6844&1590\\ \hline 
			\textbf{$T_2$}&0.92&54857&760\\ \hline 
			\textbf{$T_3$}&0.65&213369&466\\ \hline 
			\textbf{$T_4$}&0.67&376005&405\\ \hline 
			\textbf{$T_5$}&0.71&842806&346\\ \hline 
			\textbf{$T_6$}&0.65&1141010&326\\ \hline 
			\textbf{$T_7$}&0.61&1680726&303\\ \hline 
			\textbf{$T_8$}&0.47&2326448&287\\ \hline 
		\end{tabular}
		
	}
	\caption{Metric-based Topics graph layout evaluation of the Direct Approach (left) and  ZMLT (right). }
	\label{tab:tabres}
\end{table}

The trends of the average desired lengths of the computed hierarchies (the one computed by \OurAlg in orange and the one produced by the Direct Approach in gray) and the trend (in blue) of the desired edge lengths are shown in \autoref{fig:alllengthscharts}. These values have been normalized by the average length of the edges of the first level. The edge lengths are computed only on the edges of the specific levels, that is on the tree of the first level and on the forests of the others. Ideally the edge length should decrease, from the average length of the first level, by a factor proportional to the level.
While the Direct Approach tends to have the same edge length per level (except at levels 1, 3, and 8), \OurAlg shows the desired overall decrease in average edge length, which nearly matches the ideal desired edge lengths (shown in blue). This reflects the modification of ImPred made to ensure that the desired edge lengths correspond to longer edges at higher level trees and shorter edges in lower levels, as described in \autoref{sec:impred}.

\section{Limitations}
There are many limitations to the proposed approach for multi-level representation of large graph, as well as in the implementation of that approach. Here we consider a small set of limitations and briefly discuss possible directions for improvement. We use trees to obtain connected graphs for each level, although real-world road networks are not as simple. Generalizing the approach to multi-level planar subgraphs or  multi-level graph spanners (sparse subgraphs that approximately preserve shortest path distances) will likely lead to better results. 
The proposed approach also relies on  several user-defined specifications, such as the number of levels. Automatically determining the appropriate number of levels for a given graph can seem non-trivial but nevertheless needed. In the visualization we do not leverage well the possibilities of the map-based approach; for example, highways and roads still look like collections of graph edges rather than smooth curves.
Defining \emph{ highwayness} and optimizing it will likely lead to better visualizations. 



\section{Conclusions}\label{se:conclusions}
Taking inspiration from maps, our approach applies people's natural understanding of maps to a graph layout. Specifically, we presented a multi-level graph visualization algorithm that uses semantic multi-level trees of a given graph, together with the map metaphor and guarantees that the layout has the following properties: (1) appropriate representation, (2) appropriate layout, (3) real, (4) persistent, (5) overlap-free, (6) crossing-free, (7) compact. 

While the visualization relies on many existing algorithms, it provides a prototype of a multi-level graph visualization that leverages map-based representation. The seven properties provide an initial set of guidelines for such representation and our  implementation guarantees these properties. Our layout method allows for overlap removal without introducing crossings that does not rely on just scaling layout (which blows up the required area).

Code for extracting the multi-level tree, generating the layouts, measuring quality, and the interactive visualization system are available at \url{https://github.com/enggiqbal/mlgd}. The prototype is available here: \url{http://uamap-dev.arl.arizona.edu:8086/}.

\bibliographystyle{eg-alpha-doi}
\input{main-eurovis.bbl}

\end{document}

%% file: main-eurovis.bbl
\newcommand{\etalchar}[1]{$^{#1}$}